\documentclass[lettersize,journal]{IEEEtran}
\usepackage{amsmath,amsfonts}
\usepackage{algorithm,algpseudocode,algorithmicx,tabularx}
\usepackage{array}
\usepackage{multicol}
\usepackage{titlesec,color,multirow}
\usepackage{lipsum}
\usepackage{graphicx}
\usepackage{subfigure}
\usepackage{textcomp}
\usepackage{stfloats}
\usepackage{url}
\usepackage{xcolor}
\usepackage{soul}
\usepackage{hyperref}
\usepackage{verbatim}
\usepackage{graphicx}
\def\BibTeX{{\rm B\kern-.05em{\sc i\kern-.025em b}\kern-.08em
    T\kern-.1667em\lower.7ex\hbox{E}\kern-.125emX}}
\usepackage{balance}
\begin{document}
\title{Block Hunter: Federated Learning for Cyber Threat Hunting in Blockchain-based IIoT Networks}

\author{Abbas Yazdinejad, Ali Dehghantanha, \IEEEmembership{Senior Member,~IEEE}, Reza M. Parizi, \IEEEmembership{Senior Member,~IEEE}, Mohammad Hammoudeh, \IEEEmembership{Senior Member,~IEEE}, Hadis Karimipour, \IEEEmembership{Senior Member,~IEEE},  Gautam Srivastava, 
\IEEEmembership{Senior Member,~IEEE}
\thanks{A. Yazdinejad and A. Dehghantanha are with the Cyber Science Lab, University of Guelph, Canada, email: ayazdine@uoguelph.ca, adehghan@uoguelph.ca}
\thanks{R. M. Parizi is with the College of Computing and Software Engineering, Kennesaw State University, GA, USA, email: rparizi1@kennesaw.edu}
\thanks{M. Hammoudeh is with Information \& Computer Science Department, King Fahd University of Petroleum \& Minerals, Saudi Arabia
email: M.Hammoudeh@kfupm.edu.sa}
\thanks{H.Karimipour, is with the School of Engineering, Department of Electrical and Software Engineering at the University of Calgary, Alberta, Canada. email: hadis.karimipour@ucalgary.ca}
\thanks{G. Srivastava is with the Department of Math and Computer Science, Brandon University, Manitoba, Canada as well as with the Research Centre for Interneural Computing, China Medical University, Taichung, Taiwan. email: srivastavag@brandonu.ca}}
\markboth{IEEE Transactions on Industrial Informatics, 2022}%
{How to Use the IEEEtran \LaTeX \ Templates}
\maketitle
\begin{abstract}
Nowadays, blockchain-based technologies are being developed in various industries to improve data security. In the context of the Industrial Internet of Things (IIoT), a chain-based network is one of the most notable applications of blockchain technology. IIoT devices have become increasingly prevalent in our digital world, especially in support of developing smart factories. Although blockchain is a powerful tool, it is vulnerable to cyber attacks. Detecting anomalies in blockchain-based IIoT networks in smart factories is crucial in protecting networks and systems from unexpected attacks. In this paper, we use Federated Learning (FL) to build a threat hunting framework called Block Hunter to automatically hunt for attacks in blockchain-based IIoT networks. Block Hunter utilizes a cluster-based architecture for anomaly detection combined with several machine learning models in a federated environment. To the best of our knowledge, Block Hunter is the first federated threat hunting model in IIoT networks that identifies anomalous behavior while preserving privacy. \textcolor{black}{Our results prove the efficiency of the Block Hunter in detecting anomalous activities with high accuracy and minimum required bandwidth.}

\end{abstract}


\section{Introduction}
\IEEEPARstart{T}{he} 
technological trajectory of blockchain makes it a valuable tool in many areas, including healthcare, military, finance and networking, via its immutable and tamper-proof data security advantages. With the ever-increasing use of Industrial Internet of Things (IIoT) devices, the world is inevitably becoming a smarter interconnected environment; especially factories are becoming more intelligent and efficient as technology advances \cite{nn}. \textcolor{black}{ IIoT is considered a subcategory of the Internet of Things (IoT). There are, however, differences between IoT and IIoT in terms of security requirements. While the IoT makes consumers' lives easier and more convenient, the IIoT aims to increase production safety and efficiency. IIoT devices are mainly used in B2B (business-to-business) settings, while IoT devices are mostly considered in B2C (business-to-consumer) environments. This would lead to a different threat profile for IIoT networks compared to their IoT counterparts where device-to-device transactions are of utmost importance}.

IIoT networks provide an umbrella for supporting many applications and arm us to respond to users' needs, especially in an industry setting such as smart factories \cite{nn}. Blockchain technology advantages lead to its wide adoption in IIoT-based networks such as smart factories, smart homes/buildings, smart farms, smart cities, connected drones, and healthcare systems~\cite{nn, 5i}. While the focus of this paper is on the security of blockchain-based IIoT networks in smart factories~\cite{4,6}, the suggested framework may be used in other IIoT settings as well.

In modern smart factories, many devices are connected to the public networks, and many activities are supported by smart systems such as temperature monitoring systems, Internet-enabled lights, IP cameras, and IP phones. These devices \textcolor{black}{are storing private and sensitive data and may offer safety-critical services \cite{4,nn}}. \textcolor{black}{As the number of IIoT devices in smart factories increases, the main issue will be storing, collecting, and sharing data securely. Industrial, critical, and personal data are therefore at risk in such a situation. } Blockchain technology can ensure data integrity inside and outside of smart factories through strong authentication and ensure the  availability of communication backbones. Despite this, privacy and security issues are significant challenges in IIoT \cite{4,6}. \textcolor{black}{ The probability of fraudulent activity occurring in blockchain-based networks \cite{5i,6} is an important issue. Even though blockchain technology is a powerful tool, it is not protected from cyber attacks either. For example, a $51\%$ cyber-attack \cite{5i} on Ethereum Classic, and three consecutive attacks in August of 2020 \cite{55i}, which resulted in the theft of over \$5M worth of cryptocurrency, have exposed the vulnerabilities of this blockchain network.}

\textcolor{black}{ Smart factories should protect users' data privacy during transmission, usage, and storage  \cite{6}. Stored data are vulnerable to tampering by fraudsters seeking to access, alter or use the data with malicious motives.} \textcolor{black}{ Statistically speaking, these attacks can be viewed as anomalous events, exhibiting a strong deviation from usual behavior \cite{5i,7}.}
Detecting out-of-norm events are essential for threat hunting programs and protecting systems from unauthorized access by automatically identifying and filtering anomalous activities. \cite{7,8}.

The main objective of this paper is to detect suspicious users and transactions in a blockchain-based IIoT network specifically for smart factories. Here, abnormal behavior serves as a proxy for suspicious behavior as well \cite{6}. \textcolor{black}{By identifying outliers and patterns, we can leverage Machine Learning (ML) algorithms to identify out-of-norm patterns to detect attacks and anomalies on blockchain. Because deep neural networks learn representations automatically from data that they are trained on, they are the candidate solution for detecting anomalies \cite{6,8}. However, there are challenges with any ML and deep learning-based anomaly detection techniques. These methods suffer from training data scarcity problems, and privacy issues \cite{8}.}

\textcolor{black}{Detecting anomalies in the blockchain is a complicated issue \cite{10}. Not only each block needs to be sent to a central server, which increases the training time, but also the model requires new block data in the testing phase \cite{10}. In addition, when ML models are frequently updated to respond to new threats and detect anomalies, malicious adversaries can launch causative/data poisoning attacks to degrade the ML model deliberately. Attackers may intentionally send crafted payloads to evade anomaly detection.}

A novel and practical approach would be to employ Federated learning (FL) models to detect anomalies while preserving data privacy, and monitoring data quality \cite{8,fff}. FL allows edge devices to collaborate during the training stage while all data stays on the device. We can train the model on the device itself instead of sending the data to another place, and only the updates of the model are shared across the network. 

FL has become a trend in ML where smart edge devices can simultaneously develop a mutual prediction between each other \cite{8,9}. In addition, FL ensures multiple actors construct robust machine learning models without sharing data, addressing fundamental privacy, data security, and digital rights management challenges. Considering these characteristics, this paper uses an FL-based anomaly-detection framework called Block Hunter capable of detecting attack payloads in blockchain-based IIoT networks.

The main contributions of the paper are summarized as follows:  \begin{enumerate}
    \item 	Utilize a cluster-based architecture to formulate an anomaly detection problem in blockchain-based smart factories.  The cluster-based approach increase hunting efficiency in terms of bandwidth reduction and throughput in IIoT networks.
    \item	Apply a federated design model to detect anomalous behaviour in IIoT devices related to blockchain-based smart factories. This provides a privacy-preserving feature when using machine learning models in a federated framework.
    \item	Implementation of various anomaly detection algorithms such as clustering-based, statistical, subspace-based, classifier-based, and tree-based for efficient anomaly detection in smart factories.
    \item   The impact of block generation, block size, and miners on the Block Hunter framework are considered. Moreover, the performance measurements like Accuracy, Precision, Recall, F1-score, and True Positive Rate (TPR) anomaly detection are discussed.
\end{enumerate}

Here is a breakdown of the rest of the paper. Section \ref{s2} discusses anomaly detection works in the blockchain and FL. Section \ref{s3} describes the Block Hunter framework and presents the network model and topology design. In Section \ref{s4}, methodology and machine learning approaches to identify anomalies are discussed. In Section \ref{s5}, we present the assessment of the Block Hunter framework. Finally, In Section \ref{s6}, we conclude the paper and point out future work directions.

\section{Related work}\label{s2}
{\textcolor{black}{In the face of increasing cybersecurity threats and enlarging attack surfaces, it is becoming more complex and challenging to secure IIoT networks and environments \cite{n1,n2}. Furthermore, as blockchain technology is increasingly applied in a broad range of fields, anomaly detection is becoming more and more important. Anomalies can thus occur in a wide range of blockchain-based applications. This section discusses research relating to anomaly detection, especially in relation to blockchain and FL.}

In \cite{12}, the authors proposed a framework as BAD to detect anomalies in blockchain-based systems. BAD collects potential malicious activities using blockchain meta-data and has interesting features like distribution to avoid the central point of failure, trust, and privacy. 
Another work, \cite{13}, suggests blockchain and anomaly detection systems that recognize frauds when IoT meter data is tampered with. This research uses polynomial regression, DBSCAN, autoencoder, and LSTM methods to detect tampering.

The research by Sayadi \textit{et al.} \cite{14} proposes an algorithm for anomaly detection over bitcoin electronic transactions. They examined the One-Class Support Vector Machines (OCSVM) and the $K$-means algorithms to group outliers similar in both statistical significance and type. They analyzed their work by generating detection results and found that we could obtain high-performing results on accuracy.
In \cite{16}, the authors suggested an approach based on the semantics of anomalies in blockchain-based IoT Networks. A method was presented to detect anomalous behavior in blockchain that gathers metadata in forks to determine mutual informational recognition of anomalous activity. They developed a tool that improves blockchain security and connected devices. Also, in \cite{17}, has introduced encoder-decoder deep learning regression for detecting blockchain security. This work developed an anomaly detection framework that relies on aggregate information derived from bitcoin blockchain monitoring. Their experiments have demonstrated that their model can detect publicly reported attacks using the historical logs of the Ethereum network.

{\textcolor{black}{Investigation in blockchain shows blockchain Edge of Things (BEoT) can enable future services and applications, according to \cite{w1}. The authors discuss the latest developments and applications of BEoT. Their findings show that blockchain technology has grown inquisitive beyond cryptocurrency in the Edge of Things (EoT) as it provides decentralization, immutability, and traceability, in EoT systems.}

{\textcolor{black}{The field of FL is undergoing several new kinds of research.}
The article cited in \cite{8} provided the FL approach to anomaly detection in smart buildings that FL with additional recurrent neural networks is proposed as a privacy-by-design approach. It shows that it is more than twice as fast during training as its centralized counterpart. They were able to achieve superior performance in both classification and regression responsibilities compared to baseline methods. Also, in \cite{18}, Nguyen \textit{et al.} presented a self-learning federated system for detecting anomalies in IoT networks. Their system is based on device communication profiles that can detect adverse changes in IoT devices' communication. It employs FL for efficiently aggregating behavior profiles. It was one of the first systems to employ this approach to anomaly detection. Since this system can handle emerging new threats, it can be used to handle a wide variety of threats. 

The authors of \cite{19} put forward an approach via FL for detecting abnormal client behavior. The ability to detect anomalous client behaviour at the server level is mentioned in their paper. They detected abnormalities across networks using low-dimensional surrogates of model weight vectors. Experimentally, the detection-based method significantly outperforms the conventional methods based on defence. Furthermore, there is a work \cite{20} involving the use of Deep Learning and blockchain-based FL to detect COVID-19. They develop a framework to gather data from various sources and generate a global deep learning model using blockchain-based federated training. By using blockchain to authenticate the data, FL enables models to be trained while preserving privacy. By combining blockchain with federated e-learning, they developed a system for training global models collaboratively. Their results show better performance in detecting patients via this method.

 Chai \textit{et al.} \cite{21} proposed a hierarchical blockchain framework and FL to learn and share environmental data. This framework is functional and efficient for large-scale vehicular networks. FL-based learning meets the Internet of Vehicles' distributed pattern and privacy requirements. Sharing behavior is modeled as a multi-leader, multi-player trading market process to stimulate knowledge sharing. Simulated results indicate that an algorithm based on hierarchical structures can enhance sharing,  learning, and managing specific malicious attacks. Furthermore, {\textcolor{black}{the authors in \cite{w2} deliver a comprehensive investigation on how FL could supply better cybersecurity and prevent various cyberattacks in real-time. This work highlights some main challenges and future directions on which the researchers can focus for adopting FL in real-time scenarios.}

\section{Proposed Block Hunter framework in blockchain-based IIoT Networks}\label{s3}

Fig. \ref{fig01} presents a detailed overview of the proposed blockchain-based IIoT network for smart factory applications. This cluster-based architecture combines users, base stations, WiFi, service providers, and smart factories connected to the blockchain network. %to strengthen their data and keep their transactions.
Smart factories include several smart-connected devices. 
\textcolor{black}{ The service provider can collect sensor data in smart factories and use them based on their applications and services}. In addition, Fig. \ref{fig01} illustrates the relationship between the peers in terms of information between the factory and its smart devices. A transaction represents the exchange of sensitive factory information between parties during working in the blockchain network. There are several inputs and outputs in a transaction. Blocks consist of a list of transactions, a reference to the previous block, and a hash. Every block is made up of transactions that the block creator, referred to as the miner, has accepted into its memory pool from the previous block. {\textcolor{black}{Considering rigid industrial standards that should be followed when designing and implementing smart factories, it is practical to assume that the functionality of smart factories in each cluster is the same.}
\begin{figure}[!ht]%1
	\centering
	\includegraphics*[width=20pc]{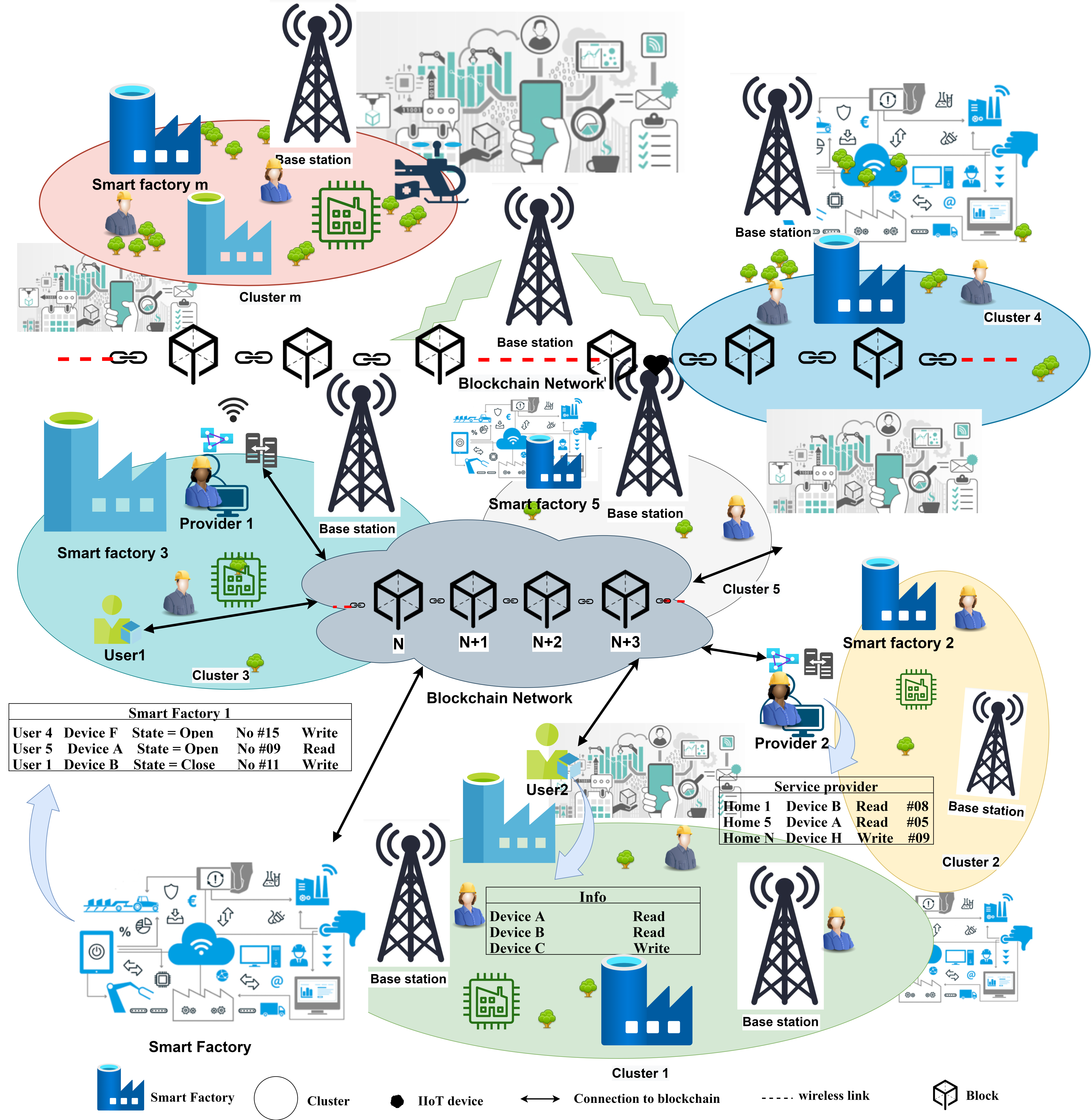}
	\caption{Overview of the blockchain-based IIoT network for smart factories}
	\label{fig01}
\end{figure}

Detecting anomalous activities is a significant contributor to automatically protecting a system from unexpected attacks. Anomalies in blockchain must be detected by sending each block of data to a central server for each block update. This is not efficient and also imposes privacy concerns. FL solutions are promising in tackling this issue. We use FL to update the model frequently and to obtain a global model for detecting an anomaly. After learning about each smart factory's data, devices, and service provider, the model's parameters will be sent to the \textcolor{black}{ parameter server} for aggregation and to update our general model.
We provide the details of implementing the Block Hunter framework in the following sub-sections.

\subsection{Role of FL to detect anomalies in Block Hunter }
We distribute local models across the blockchain-based IIoT network instead of learning an anomaly detection model and evaluating it on a single node. As shown in Fig. \ref{fig02}, the FL setup involves local models as well as distributed smart factories nodes. Instead of a centralized learning environment, $K$ smart factories learn a local model in an FL manner. {\textcolor{black}{ The $k = 1, \ldots, K$ local models have the same structure, but they are trained with different datasets that originate from their connected clients. Our proposed federated anomaly-detection algorithm for smart factories in IIoT networks is shown in Algorithm \ref{aa}. $C$ represents the batch size for the global operation; $B$ determines the local batch's size; a factor of $k$ indexes the $K$ smart factories; $E$ indicates the number of local epochs, and $h$ represents the rate of learning.} Initializing begins the process of initializing the model parameters. During the training step, \texttt{FedAvg} \cite{22} is sent to smart factories and updates the model. Finally, our updated trained model can be tested to detect any anomalies.

\begin{figure}[!ht]%1
	\centering
	\includegraphics*[scale=0.029]{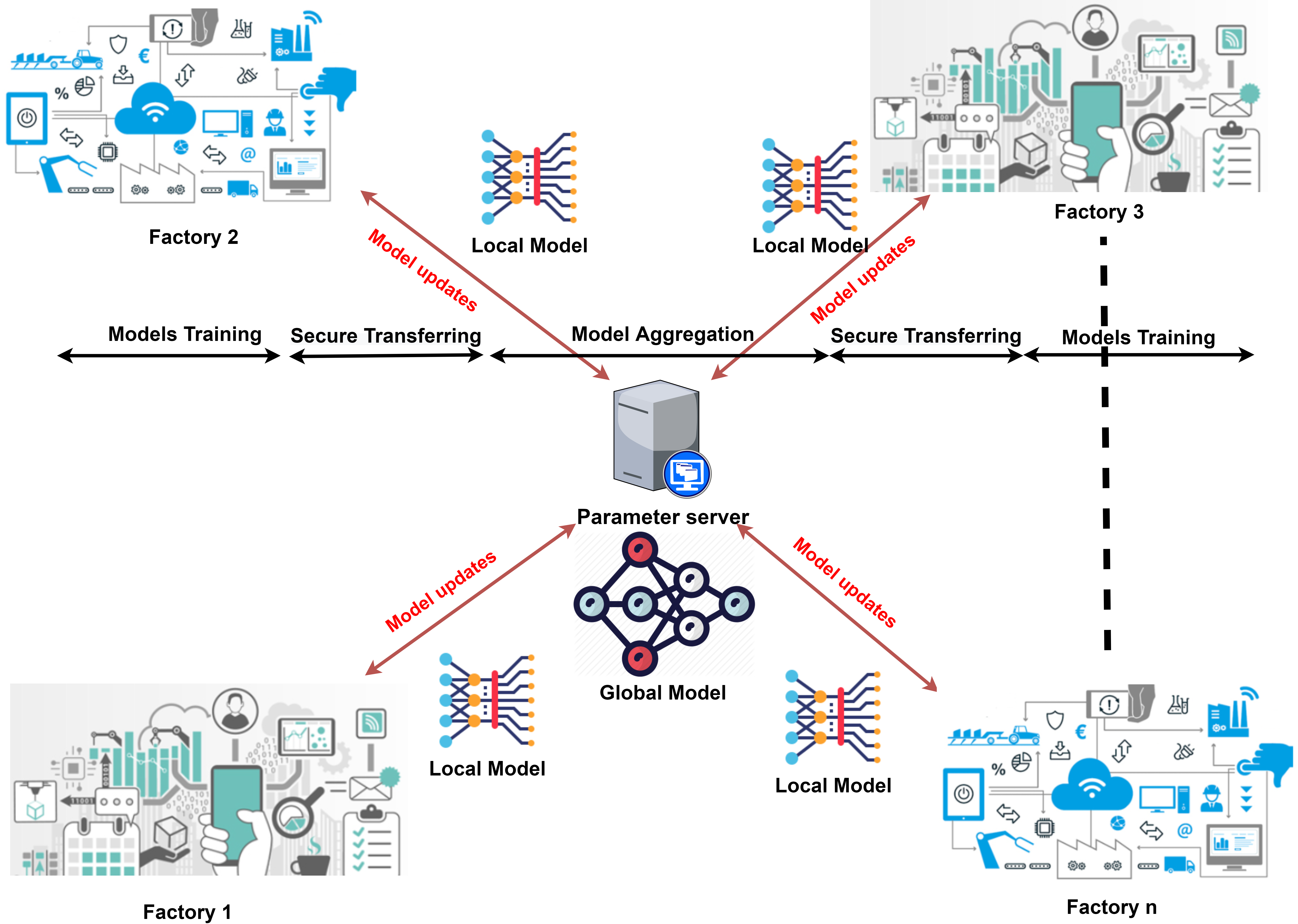}
	\caption{Proposed Federated Anomaly-detection in Block Hunter framework}
	\label{fig02}
\end{figure}

Based on Algorithm \ref{aa}, the parameter server starts the FL scheme at $t = 0$, initializing local models with the first set of weights. Next, these local models are downloaded from the parameter server to each of the $k = 1, \ldots , K$ smart factory. Third, using the corresponding blockchain datasets' training data, each of the $E = 1, \ldots , E$ local models computes in parallel a new local set of weights. Finally, the parameter server aggregates the weights in each client's local model to create an improved global model using weighted averaging (weighted average). Each time the cycle repeats, a new epoch is initiated until a certain stop criterion has been reached. According to McMahan, we utilize the same design policy for FL as he does \cite{22,23}. They develop the FL problem as a federated optimization problem by distributing the model, $m\_t$ to (a subset of) $K$ clients of the learning federation at time ${t}_{0}$. 
To summarize, steps for FL in the Block Hunter framework are categorized:
\begin{itemize}
    \item  \textbf{Federation Construction:} The subset of smart factory members, cluster, selected to receive the model locally.
    \item  \textbf{Decentralized Training:} When a cluster of smart factories is selected, it updates its model using its local data.
    \item  \textbf{Model Accumulation: }\textcolor{black}{ Responsible for accumulating and merging the data models. 
Data is not sent and integrated from the federation to the server individually.}
    \item  \textbf{Model Aggregation (\texttt{FedAvg}):} Parameter server aggregates model weights to compute an enhanced global model.
\end{itemize}

\begin{algorithm}[!ht]
	\tiny
	\caption{\small {Federated anomaly-detection model} }
	\tiny
	\begin{algorithmic}[1]
		%\State  \textbf{Parameter}: q, W, F, y, U, Ni, X, t
			\State   \textbf{{Input:}} {Pre-trained model}
				\State   \textbf{{Output:}}  {Global anomaly-detection model}
			\State   \textbf{Initializing:  ($t = 0$)} // {\textcolor{black}{ start to set values}}
			\State \qquad  Parameter server update ($W$)  
		\State \qquad   Define initial values = ${B, E, h, C, K}$
		\State \qquad  \qquad  initial local models ()
		\State \qquad    Model (${m}_{i}$) = set parameters ($w_{1}, \ldots , w_{n}$)
		\State \qquad   Client model update 

     	\State \qquad   Local models$\mathrm{\to}$ smart factories' clusters
      
			\State   \textbf{Federated Training:} // {\textcolor{black}{ beginning FL method }}
	\State \qquad \qquad    While ($K > 0$)
		\State \qquad \qquad  \qquad Get (local models)
			\State \qquad \qquad \qquad For local epoch $E$
				\State \qquad \qquad   \qquad \qquad    For each batch $1$ to $b$
					\State \qquad \qquad \qquad \qquad \qquad    Run local models
						\State \qquad \qquad \qquad  \qquad Obtain and set model parameters
							\State \qquad \qquad  \qquad Return (Model parameters)
						
							\State \qquad \qquad  \qquad Parameter server =  \texttt{FedAvg} ($w$)
						\State	\qquad	 \textbf{		Update ($w, m$)}
	
	\State  \quad   \quad \quad Decrypted ($w$,$m$) 
	\State  \quad  \quad  \quad $D = $ split Data into batches of size $B$
	\State  \quad   \quad \quad \quad for local epoch $E$ from $1$ to $B$ do
	\State  \quad   \quad \quad \quad  \quad \quad	for batch $x$ $<=$ $B$ do
	\State  \quad   \quad \quad return $w$ and $D$ to server
     		\State \qquad Return updated Model

	\State   \textbf{{FedAvg}: } {\textcolor{black}{// model aggregation}}
		\State \qquad \qquad Server ( Initialize ${w}_{0}$)
		\State \qquad \qquad  encryption (Homomorphic)
		\State \qquad \qquad \qquad $P$ = compute  loss ($w$, $b$)
		\State \qquad \qquad for  round $c = 1$ to $k$ do
		\State \qquad \qquad \qquad \qquad Server: Send $w_{t-1}$ to Smart factory $i-1$
		\State \qquad \qquad \qquad\qquad $E = E + 1$
		\State \qquad \qquad \qquad Parameter server = Update ($w_{k-1}, \ldots , w_{k}$)

	\end{algorithmic}
	\label{aa}
\end{algorithm}

{\textcolor{black}{ At runtime, pre-trained models as local models are sent to clusters in the Block Hunter framework from the Parameter server, considering blockchain-based IIoT networks environments. The local models are sent to smart factories for training based on local epochs. Then the parameters and hyperparameters will be forwarded to the Parameter Server for aggregating model weights to compute the global model. The global model is an ML model that holds in the Parameter Server to update its parameters. } {\textcolor{black}{ When a new cluster joins our framework, the latest global model will send to that cluster as a pre-train model that in the real-world application, we can simply follow this approach. The updated global model is sent to clusters gradually during evolution.}

\subsection{Anomaly Detection in  blockchain-based IIoT network}
Smart factories have sensitive data, so storing it on the blockchain with its limited storage is both financially and computationally costly. Therefore, the actual smart device and sensor data are stored in the smart factory.  %A smart home handles data and keeps a transaction in the block of the data to verify the data owner.  
The smart factory data includes information about the type of data and control states as well.

The premise behind the development of an anomaly detection framework for the blockchain-based IIoT networks in smart factories lies in providing a new decentralized system based on FL that leverages all smart factory data while protecting their privacy. Additionally, we will reach a point where we need to attend to the issue of a fork in the blockchain scope during anomaly detection. In some instances, devices or nodes do not agree on the state of the blockchain, leading to the fork issue in the blockchain. When blockchain-based applications are being developed, forks become more concerned because they have the potential to be used for malicious purposes. Indeed, a global ML model can use all of the collected information from previous forks to detect anomalies during training. This approach has the advantage that while attacks may only happen once within a smart factory, they behave the same way when repeated against other smart factories over time. Hence, information on past attacks may help us blacklist them and prevent them from occurring in the future. The advantage of FL is clear since it will train the global ML model for anomaly detection.

\textcolor{black}{Based on Fig. \ref{fig01}, each participant in a smart factory can provide a fake blockchain transaction as a side-channel to deliver a message.} A malicious transaction, as well as creating a fake block, are also possible in this situation. A malicious transaction is a special type of fake transaction, which consists of a hidden message which is aimed at disrupting the network by hitting a specific peer. \textcolor{black}{ Inserting} fake blocks are blockchain blocks that contain one or more stolen/malicious transactions. Fake blocks can either be eventually discarded or accepted into the mainstream chain.

Our solution considers smart factories' data and chain forks. We collect, enrich, and share such information with other local ML models across the network. We used the specific information for training anomaly detection in each local ML model that contains sensitive smart factory data, the features of previous forks, and the number and type of malicious transactions that occurred. As a result, we can hunt an anomaly by Block Hunter in a blockchain-based IIoT network for smart factories. To protect the privacy of the data, we only share the parameters of trained models instead of the original data from smart factories and their blockchain. This work aims to train a global anomalous detection model through locally trained sub-protocol models based on the Block Hunter framework. 

Regarding the threat model, the solution proposed in this paper has been designed to be resilient against any class of attacks where a malicious entity can append to the blockchain system.

\subsection{Network model and topology design }
This section discusses the efficient network model and topology design for blockchain-based IIoT networks. Wireless sensor networks have a variety of topologies, which affect their performance and behavior. Some of the metrics include throughput, reliability, energy consumption, and latency \cite{24}. Therefore, we propose blockchain technology's cluster-based formation model for smart factories. Cluster-based architecture provides more efficient use of resources \cite{a4} and throughput during the blockchain run in each smart factory. Clustering reduces the computational complexity in the creation of the underlying network through a hierarchical approach \cite{24}. It is especially so with \textcolor{black}{blockchain-based IIoT networks }that are expected to encompass large numbers of individual devices. Also, we believe that cluster-based architecture will enable us to hunt and manage anomalies better in each smart factory zone and increase the network's throughput.

In each cluster, the smart factory controls all IIoT devices' activities, and one of the smart factories is usually known as Cluster Head (CH) or a leader node. A CH can perform extra duties in blockchain-based networks like taking part in the mining process by reviewing aspects such as energy, memory, and computing power. Fig. \ref{fig01} shows the clustering strategy in the blockchain-based smart factory network model.

Based on the target Block Hunter framework, which can be represented as a directed graph  $G= (S, D)$ with $ D $ being the set of IIoT devices, representing smart devices, $D=\{{d}_{1}, {d}_{2}, \ldots , {d}_{n}\}$,  and $S=\{s_{1}, s_{2}, \ldots , s_{n}\}$ is the set of smart factories in each cluster.  In ${S}_{1T}=[{t}_{1}, {t}_{2}, \ldots , {t}_{m}]$, we consider the set of transactions in smart factory ${S}_{1}$ that belongs to the blockchain network. $B = [b_{1},b_{2}, \ldots ,b_{3}]$ represents the number of existing blocks in the blockchain network. More formally, ${s}_{n}=\bigcup _{j}^{k} {s}_{j} \times {D}_{kj} $, with $k$ being the number of deployed clusters, and $j$ is smart factories in that cluster. It should be noted that the set of IoT devices refer to a smart factory, $D_{kj}\in [1,j]$ in the $K^{th}$ cluster.

It is possible to summarize the distribution of smart factories with their devices in the Block Hunter, the proposed cluster-based architecture, in Equation 1, distance function, \textit{Df}. This is the point at which smart factories and IoT devices will cluster based on most centrality derived from a distance measure based on the presence or absence of shared neighboring devices in the space of ($i,j$).

\begin{equation}\label{eq1}\small
D_{f} (D_{ki},D_{kj})= \sqrt{\sum_{i,j=1}^{K} (S_{nj}-S_{ni})^2\times(D_{kj}-D_{ki})^2}
\end{equation}
The clustering part is shown in Algorithm \ref{a2}, and it can be considered a piece of the overall algorithm in the Block Hunter. Algorithm \ref{a2} collects the locations of smart factories and their IIoT devices. Based on Equation \ref{eq1}, we measure each smart factories' distance and their devices and record it until we obtain the cluster-based architecture. Afterward, the cluster calculates a collection of $S$ nearest smart factories for each IoT device, $D$.

\begin{algorithm}
	\tiny
	\caption{\small {\textcolor{black}{Cluster formation strategy in Block Hunter} }}
	\tiny
	\begin{algorithmic}[1]
		%\State  \textbf{Parameter}: q, W, F, y, U, Ni, X, t
		\State   \textbf{Input:}
      \State \qquad Get = ($S, D$) 
\State   \textbf{Initialize:}
        \State \qquad Define initial values
       \State \qquad Set Number of Cluster = K
       \State \qquad Get loc =  $S=\{s_{1}, s_{2}, \ldots , s_{n}\}$ {\textcolor{black}{// location of smart factories}}
       \State \qquad Get loc = $D=\{d_{1}, d_{2}, \ldots , d_{n}\}$ {\textcolor{black}{// location of devices}}
       \State \qquad ${s}_{n}=\bigcup _{j}^{k} {s}_{j} \times {D}_{kj} $ //{\textcolor{black}{ Deployed clusters and smart factories}}

\State   \textbf{Main():}
 	\State \qquad \textbf{Get ($K, S, D$)}
		%\State \qquad  Parsing incoming packets base on parse graph, Return extracted fields:
 \State \qquad \textbf{While( $K$ $>$ 0)} 
 \State \qquad 	$\{$
		\State \qquad \qquad 	\textbf{ For $z$ each $K$}
      \State   \qquad   \qquad \qquad Comput = $D_{fz}$ 
         \State     \qquad    \qquad   \qquad \qquad ($D_{ki},D_{kj}$) =  $\sqrt{\sum_{i,j=1}^{K} (S_{nj}-S_{ni})^2\times(D_{kj}-D_{ki})^2}$ //{\textcolor{black}{ Calculating distance for finding neighboring devices. The presence or absence of devices in the space of (i, j)}}
      
		 \State     \qquad   \qquad \qquad Set\_area = $K_{z}$
		  \State  \qquad \qquad Client distance update 
		 \State \qquad \qquad  Marge neighbor
		
	 \State \qquad 	$\}$
	 \State	\textbf{	Return $K$}
	
	\end{algorithmic}\label{a2}
\end{algorithm}
{\textcolor{black}{Setting model parameters in the parameter server and sending pre-trained models to clusters happen during initializing. Next, local models are trained by clusters in the training step to aggregate models and update the global model parameters. }

\section{Methodology}\label{s4}
In this section, we study several machine learning techniques for identifying and detecting anomalies in the Block Hunter framework. 

\subsection{Neural Encoder-Decoder (NED) model}
An example of a classifier-based anomaly detection algorithm is the neural encoder-decoder model. The proposed anomaly detection framework develops a neural encoder-decoder model that summarizes the information about the blockchain's status and transactions and then rebuilds the initial data from this space. Encoding/decoding preserves the data's basic properties when the current status is consistent. Differently, anomalous situations exhibit inconsistent values, ultimately leading to a failing reconstruction. In an encoder-decoder, this quantity would be paraphrased as noise and therefore would be failing when reconstructing. Therefore, the difference between the initial and reconstructed values would highlight the anomalous and abnormal situation, thereby triggering an alert \cite{10,27}.
Neural encoder-decoder models analyze sequences of temporally sorted events. In general, we suppose that the data will be sequenced as $P = \{P_{1}, P_{2},\ldots , P_{n}\} $ concerning some period of observation, where $P_{t}$ is an assessment of the properties of the $t^{th}$ event in the chronological order of events in $P$. Anomalous events occur in $P$, i.e. a vector ${P}_{t}$ drastically various from its neighbors ${P}_{t}$.

\subsection{Isolation Forest}	
The Isolation Forest (IF) model falls under the Tree-based anomaly detection algorithms category. The approach has gained much universal acceptance in recent years because it is unsupervised. Isolation forest is a concept based on the idea that it's more prudent to isolate data anomalies rather than generalize the norms. It's a recursive and random partitioning process to isolate the anomalous data point in the dataset until it simply describes the stored data. A tree structure represents the recursive partition. A forest of isolation trees is the foundation of the isolation forest algorithm, where cells in the dataset are randomly selected from the data to form a forest of normal and outlier cells. These trees are binary trees that have zero or two child nodes, and an isolation forest contains isolation trees of this type \cite{10,28}. Consider that $X$ is either a leaf node that does not have any children or a parent node that has two children named $XL$ and $XR$. To choose which child nodes belong to which parents, a test must be attached to node $X$. The testing procedure involves selecting a random feature f across all the data points and an arbitrary splitting point $q$. Node f $< $ q is in the zone of $XL$, and $f \geq  q$ is in the zone of $XR$.

\subsection{Cluster-Based Local Outlier Factor}
Our cluster-based local outlier factor (CBLOF) model belongs to the classifier-based algorithm-based anomaly detection category. Our cluster-based local outlier factor (CBLOF) model belongs to the classifier-based algorithm-based anomaly detection category. Within this algorithm's anomaly detection methodology, the data is clustered into clusters, based on which anomaly scores can be computed similar to those of the local outlier factor algorithm and so on. This algorithm's underlying principle of anomaly detection is based on clustering data sets together. This algorithm creates clusters using groups in a dataset by arbitrary clustering algorithms that assign a specific observation to a cluster. The clusters are sorted in each case corresponding to their respective sizes of $|{F}_{1}| > |{F}_{2}| >\ldots > |{F}_{k}|$  where ${F}_{1}, {F}_{2}, \ldots , {F}_{k}$ all represent the cluster for which number $k$ is the cluster number \cite{10}.
A pair of clusters, when intersecting with each other, should give rise to an empty set. However, all these clusters' unions should represent all of the observations in Dataset $D$. we are supposed to search for a boundary index value that separates the Small Clusters from the Large Clusters. Finally, we calculate the CBLOF scores for each observation by using the following equation, $1\le i\le k$:
\begin{equation}\label{eq2}\small
CBLOF (t)=\begin{cases}
|Fk|.dis(t,Fi)  & \mathrm {	t\in Fi }\\
|Fk|.min(dist(t,Fi)      & \mathrm{  t\in Fi , } 
\end{cases}
\end{equation}

\subsection{Principal Component Analysis (PCA)}
A PCA model is a subsequence-based anomaly detection algorithm. PCA is commonly considered a method to reduce the dimensionality algorithm. The variance-covariance of dataset characteristics can be used to construct new variables known as principal components, which are functions of original variables. For principal component analysis, one uses p distinct linear combinations of random variables ${x}_{1}, {x}_{2}, \ldots , {x}_{p}$. The principal component has the following characteristics: they are uncorrelated to each other. Each component's variance decreases in descending order, with the principal component containing the highest variance and the subsequent details having lower variances. When combining all the principal components' variations, the sum of the total variation of the original features is always equal to the total variation of all the principal components. To estimate the principal components of a system, we can use eigenanalysis to get the correlation matrix, or covariance matrix of data features \cite{10,30}.

The PCA algorithm detects anomalies by getting rid of any outliers. The outliers are determined by Mahalanobis distance that is carried out repeatedly to eliminate all data points with high Mahalanobis distance values. Where $S$ is a covariance matrix, ${x}_{i}$ is the measure of an observation of the $i^{th}$ feature in data, and $x$ is the mean of all observations, Mahalanobis distance is denoted as follows:

\begin{equation}\label{eq3}\small
D = \sqrt{{\left(Xi - X\right)}^{T}{s}^{-1}{\left(Xi - X\right)}}
\end{equation}

\subsection{$K$-means}
In the cluster-based detection algorithms category, $K$-means is a clustering-based algorithm.  As one of the most popular clustering algorithms, $K$-means is also commonly used as an anomaly detection algorithm. It has been introduced as an unsupervised learning scheme. The data is divided into $k$ different clusters, with each sample belonging to the cluster with the closest mean value within each cluster. Across clusters, there is a cluster centroid $c$, which is the mean of observations from each cluster in that cluster. When assessing the similarities among independent observations, the similarity measure employed is Euclidean distance, where ${x}_{i}$ is the measurements and ${c}_{i}$ is the centroids, and n outlines the number of independent measurements \cite{10,31}.

\begin{equation}\label{eq4}\small
{d}^{2}\left(X , c\right)= \sum _{i=1}^{k}{\left(Xi-ci\right)}^{2}
\end{equation}

\section{Discussion \& Evaluation}\label{s5}
This section evaluates the performance of the Block Hunter and provides results and discussion.

\subsection{Experimental Setup}
We formed an experimental setup on Intel(R) Core(TM) i7-10700KF CPU @ 3.80GHz   3.79 GHz, Linux 64-bit operating system (Ubuntu 20.04), and equipped it with 16 GB DDR4 memory. To evaluate our network model and cluster-based topology design in the proposed framework, we apply Bitcoin Simulator\footnote{\url{https://github.com/ctch3ng/Bitcoin-Simulator-NS3}}.  Bitcoin Simulator is an open-source bitcoin simulator developed on NS3. The Bitcoin Simulator has been tested with NS3. Also, we consider LENA as an NS-3 module to simulate 3GPP networks. The NR module is a pluggable module for NS-3 that can be used to simulate New Radio (NR) cellular networks. %The simulator is the natural evolution of LENA, the LTE/EPC network simulator. 5G-LENA features are pulled from the NSNAM application website\footnote{\url{https://apps.nsnam.org/app/nr/}}}.
NS-3 supports the widest variety of network models and protocols and supports the greatest variety of networking devices. Indeed, wireless networks and protocols rely on the NS-3 to determine their performance. Therefore, the assessment of the proposed federated framework will do on the performance metrics such as the impact of block size, number of blocks, number of IoT devices, and number of miners.
The implementation details of the Block Hunter framework  \textcolor{black}{is presented in Table \ref{t1}. }

\begin{table}[!ht]
\tiny
\centering
	\caption{Federated framework parameters}
\begin{tabular}{|c|c|}
\hline
\textbf{Parameters}          & \textbf{Description}          \\ \hline
\textcolor{black}{ Simulators }                  &\textcolor{black}{ Bitcoin simulator/ NS3 / 5G-LENA }\\ \hline
Operating system             & Ubuntu 20.04           \\ \hline
libraries                    & PySyft / Pythorch             \\ \hline
Number of Clusters           & $50$                            \\ \hline
Optimization method          & SGD                           \\ \hline
local epoch         &  E = 4      \\ \hline
 Fraction of smart factories     &\ $6e- 3$                     \\ \hline

\textcolor{black}{Mobility model }     &\textcolor{black}{ Random waypoint model          }               \\ \hline
\textcolor{black}{Traffic Type  }     &\textcolor{black}{Constant Bit Rate (CBR)        } 
 \\ \hline

Number of IoT devices        & $5000$                          \\ \hline
Block Size                   & $1$ MB, $2$ MB, $4$ MB, $8$ MB, $16$ MB    \\ \hline
Number of Miners              & $16, 32, 64$                    \\ \hline
Local epochs in each cluster & $4$                             \\ \hline
Learning rate                & $3e- 2$                         \\ \hline
Local mini-batch size        & $15$                            \\ \hline
Federated approach           & \texttt{FedAvg}   
\\  \hline

\end{tabular}
\label{t1}
\\
\end{table}
For the federated setup, we have considered PySyft\footnote{\url{https://github.com/OpenMined/PySyft}} and PyTorch\footnote{\url{https://pytorch.org}}. PySyft is an open-source library that allows us to create VirtualWorkers for training our machine learning models to detect an anomaly. It is designed to allow users to create a private and secure ML model, and it is built into some existing ML libraries, such as PyTorch. Our framework is trained with the \texttt{FedAvg} method with $E = 4$ local epoch and fraction $c = 6e -3$. \textcolor{black}{ $E$ mentions to Local batch size used at each learning iteration, and $c$ refers to the number of smart factories used at each iteration. It is also important to emphasize $e$. The $e$ is denoted $exp$, which is short for $exponential$. } Finally, an SGD optimizer is used for training the models with a learning rate of $3e - 2$.
\begin{figure*}
\centering
  \begin{minipage}{.5\linewidth}
    \makebox[.6\linewidth]{\includegraphics[width=.65\linewidth]{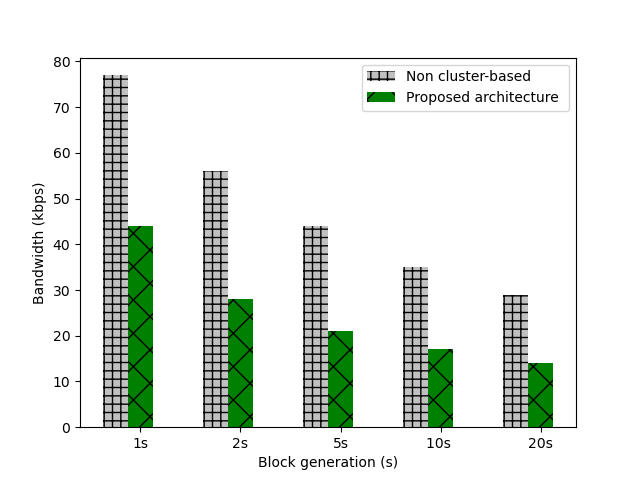}}%
    \makebox[.6\linewidth]{\includegraphics[width=.65\linewidth]{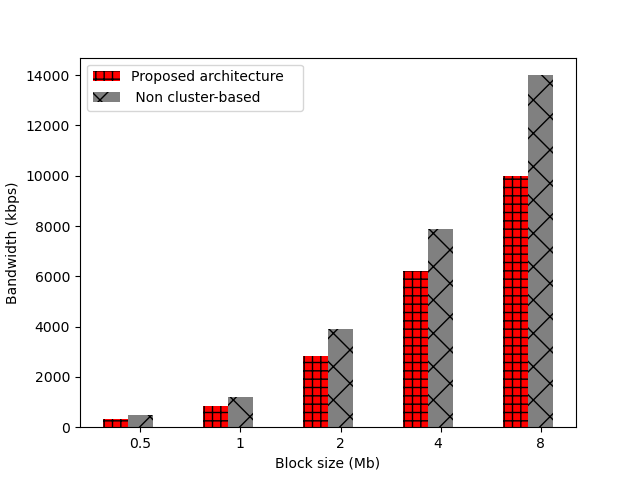}}

    \makebox[.6\linewidth]{\tiny (a) Impact of the Block generation on bandwidth consumption}%
    \makebox[.6\linewidth]{\tiny (b) Impact of the Block size on bandwidth}%

    \medskip

    \makebox[.6\linewidth]{\includegraphics[width=.65\linewidth]{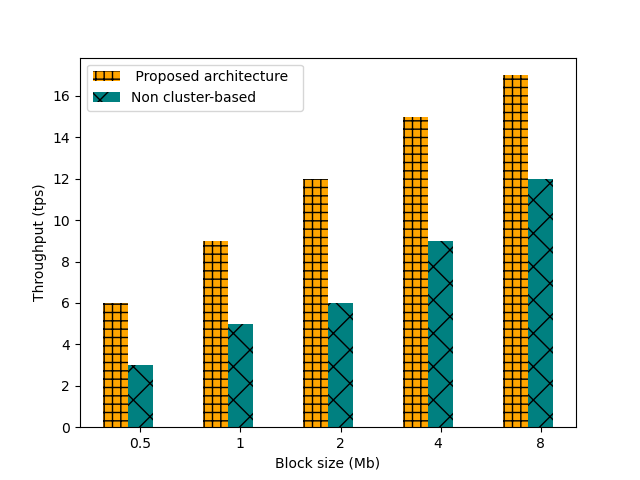}}%
      \makebox[.6\linewidth]{\includegraphics[width=.65\linewidth]{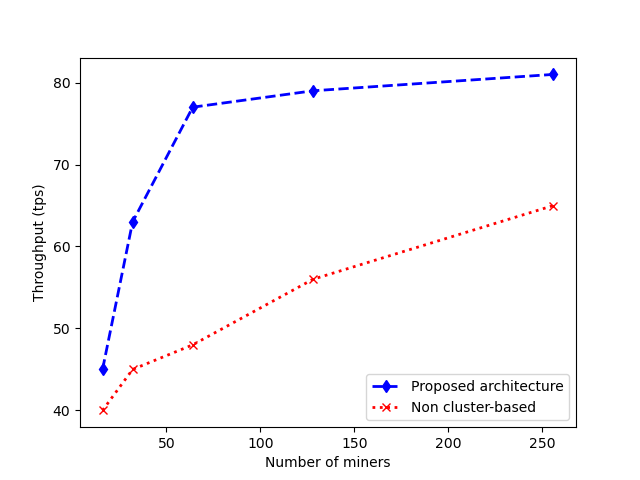}}

    \makebox[.6\linewidth]{\tiny (c) Impact of the Block size on throughput}%
    \makebox[.6\linewidth]{\tiny (d) Impact of number of miners}%
  \end{minipage}%
      \caption{Cluster-based architecture evaluation}
    \label{fig:sample_subfigures}
\end{figure*}

\begin{comment}
\begin{figure}
    \centering
    \subfigure[Impact of the Block generation on bandwidth consumption]
    {
        \includegraphics[width=2in]{7n.png}
        \label{first_sub}
    }
    
    \subfigure[Impact of the Block size on bandwidth]
    {\includegraphics[width=2in]{8n.png}
        \label{second_sub}
    }\hfill
    \subfigure[Impact of the Block size on throughput]
    {\includegraphics[width=2in]{9n.png}
        \label{third_sub}
    }
    
    \subfigure[Impact of number of miners]
    {\includegraphics[width=2in]{10n.png}
        \label{fig:forth_sub}
    }
    \caption{Cluster-based architecture evaluation}
    \label{fig:sample_subfigures}
\end{figure}
\end{comment}

\subsection{Datasets}
{\textcolor{black}{The proposed framework is evaluated by two datasets on the blockchain side, providing conditions for blockchain adoption in smart manufacturing systems, and also two IIoT-related datasets for assessing BlockHunter for smart factories.} 

The Bitcoin Transaction Dataset (BTD)\footnote{\url{https://ieee-dataport.org/open-access/bitcoin-transactions-data-2011-2013}} designed for research on blockchain anomaly and fraud detection. It has been donated to the IEEE data port online community for academic exploration. Because the dataset is imbalanced and contains roughly $30$ million transactions, it presents a challenge in creating an anomaly-detection model that captures all of them. The dataset is an implementation of a research project that presents anomaly detection within the context of blockchain technology and its applications in the monetary domain. It extracts blockchain data and uses machine learning techniques to hunt potentially malicious transactions.

Another dataset is Ethereum Classic (ETC)\footnote{\url{https://www.kaggle.com/bigquery/crypto-ethereum-classic}} that is a BigQuery Dataset. We will be able to access Ethereum Classic transactions and block history in this dataset. The Ethereum Classic open-source, based on the Ethereum platform, is a platform that enables distributed computing by using a public, distributed decentralized network for executing scripts with the ability to manage smart contracts. The dataset consists of all blocks, contracts, logs, tokens, traces, and transactions contained within the blockchain network.

{\textcolor{black}{In choosing IIoT related datasets, two well-known datasets have been considered: Gas Pipeline (GP) and Secure Water Treatment (SWaT). They are well fit for the IIoT environments and are publicly available \cite{gas,sw}.}

\subsection{Experimental Analysis}
%\subsubsection
A cluster-based architecture provides more efficient use of resources and throughput during the blockchain run in smart factory applications.
To evaluate the performance of the Block Hunter, cluster-based architecture, the simulation parameters are presented in Table \ref{t1}.To accomplish more realistic results, we did the simulation $20$ times and designed another scenario as {\textcolor{black}{ a non-cluster model to compare the architectural models' performance during the simulation. The non-cluster model combined blockchain technology with the standard network model and did not consider and divide it into cluster architecture. It has no features and typologies of cluster-based architecture such as adjacencies with other clusters or part of the network, flexibility, and scalability during run time. Conversely, in cluster-based architecture, each cluster has adjacencies with other clusters and supports the dynamic characteristics of a network.} %The thing that needs to be highlighted here is that these parameters, such as block size, number of blocks, nodes, and relay network, make this evaluation reproducible.
In the following, we address the impact of Block generation, the impact of the Block size, and the impact of the number of miners in the evaluation. {\textcolor{black}{ In the proposed framework, the public blockchain network is deployed among clusters that include smart factories. We need a public blockchain to allow any smart factories to join and keep the system completely decentralized. Additionally, public blockchains give all participants equal access to the chain.}

\begin{itemize}
 \item \textbf{Impact of the Block generation}\\
Block generation interval is regarded as an important metric for measuring the performance of blockchain networks. If we have an organized topology and structure, the block generation will be more efficient to support nodes than in a distributed network with a solid and organized structure. Further, since blocks are generated more frequently in individual clusters instead of generated in batches that consume a considerable amount of bandwidth, we can better manage and use the bandwidth. Fig. 3(a) shows the bandwidth efficiency of the cluster-based design (proposed architecture) compared to the non-cluster-based design.

%Fig. \ref{first_sub}

Although there is an increase in block generation time in the non-cluster-based design, more blocks will be generated in the network and consume more and more bandwidth. Hence, cluster-based architecture provides better performance since the nodes are distributed across the whole network (currently, there are $5000$ reachable nodes in $K = 50$ Clusters).

 \item \textbf{Impact of the Block size}\\
Block size has a significant impact on the performance of blockchain. The block size determines the highest number of transactions that can be approved within a block. This size, thus, controls the throughput (transactions/second) obtained by the proposed design. Larger blocks cause more sluggish propagation in each cluster than smaller blocks. In Fig. 3(b) and Fig. 3(c), we show that the bandwidth consumption and throughput increase with increasing block size from 0.5 MB to 8 MB. This directly impacts both the bandwidth and throughput of the proposed model. As expected, Block Hunter has a higher performance because of better network communication, efficient topology management, and minimized delay.
% \ref{second_sub}  \ref{third_sub} \ref{fig:forth_sub}
 \item \textbf{Impact of number of miners}
The number of miners in a given architecture directly impacts throughput (transactions/second). According to Fig. 3(d), an increasing number of miners from $16$ to $256$ and the block size to $1$ MB in all clusters increased the model throughput. The increase in the number of miners makes it easier for smart factories to reach a consensus. Additionally, the proposed cluster-based architecture can handle more transactions in each block by increasing the block size. Consequently, it will grow the proposed architecture's throughput rate and offer better performance.
\end{itemize}

\begin{figure*}
  \centering
  \begin{minipage}{.6\linewidth}
    \makebox[.6\linewidth]{\includegraphics[width=.51\linewidth]{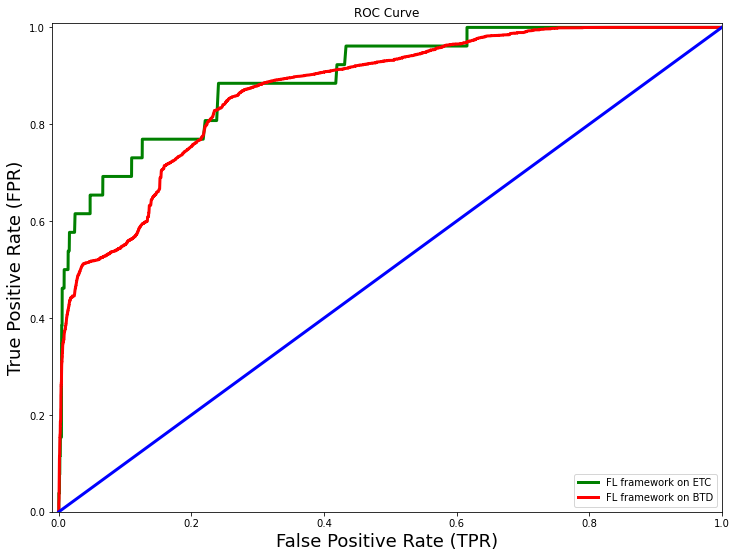}}%
    \makebox[.6\linewidth]{\includegraphics[width=.51\linewidth]{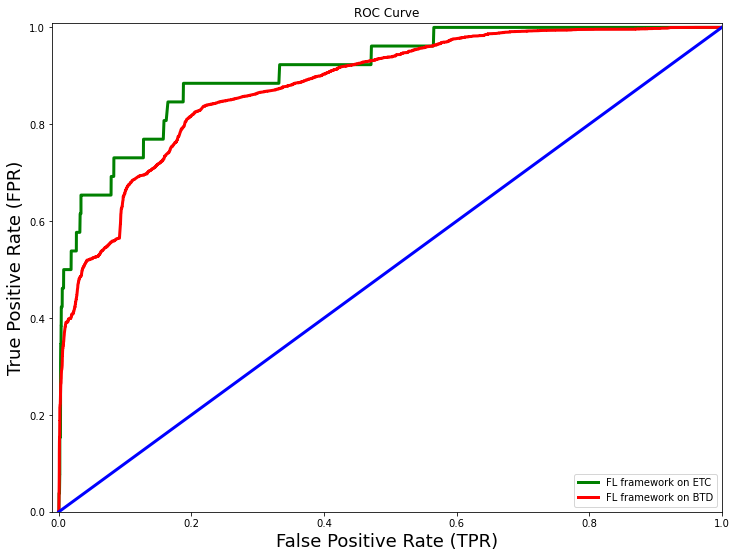}}

    \makebox[.6\linewidth]{\tiny (a) The AUC of ROC curves for CBLOF}%
    \makebox[.6\linewidth]{\tiny (b) The AUC of ROC curves for $K$-means}%

    \medskip

    \makebox[.6\linewidth]{\includegraphics[width=.61\linewidth]{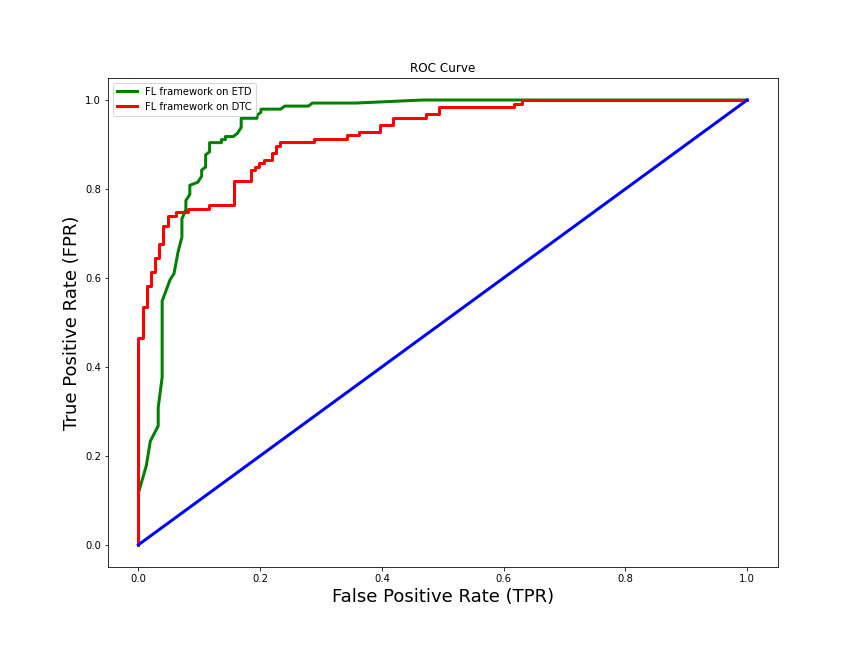}}%
    \makebox[.6\linewidth]{\includegraphics[width=.61\linewidth]{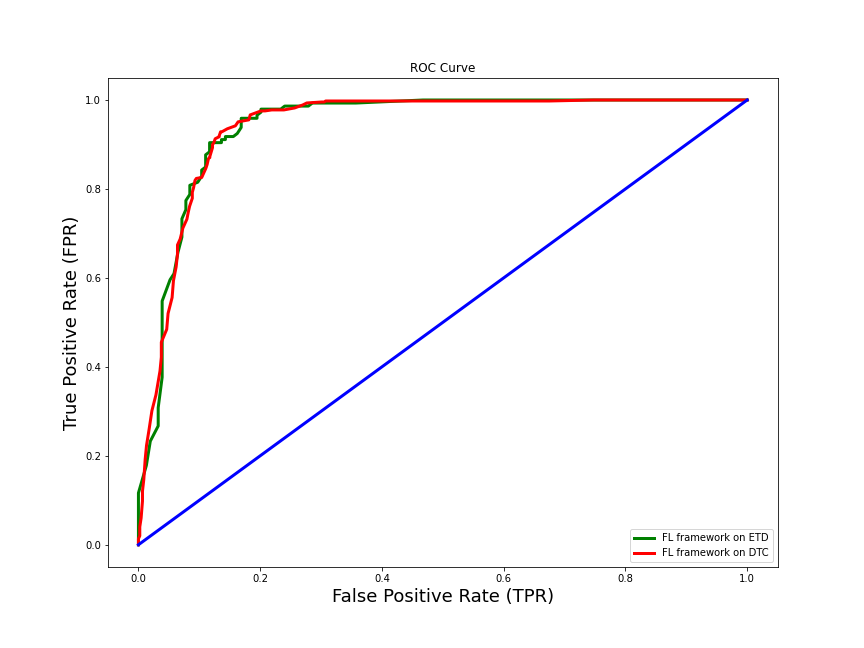}}

    \makebox[.6\linewidth]{\tiny (c) The AUC of ROC curves for PCA}%
    \makebox[.6\linewidth]{\tiny (d) The AUC of ROC curves for IF}%
  \end{minipage}%
  \begin{minipage}{.5\linewidth}
    \centering
    \includegraphics[width=.72\linewidth]{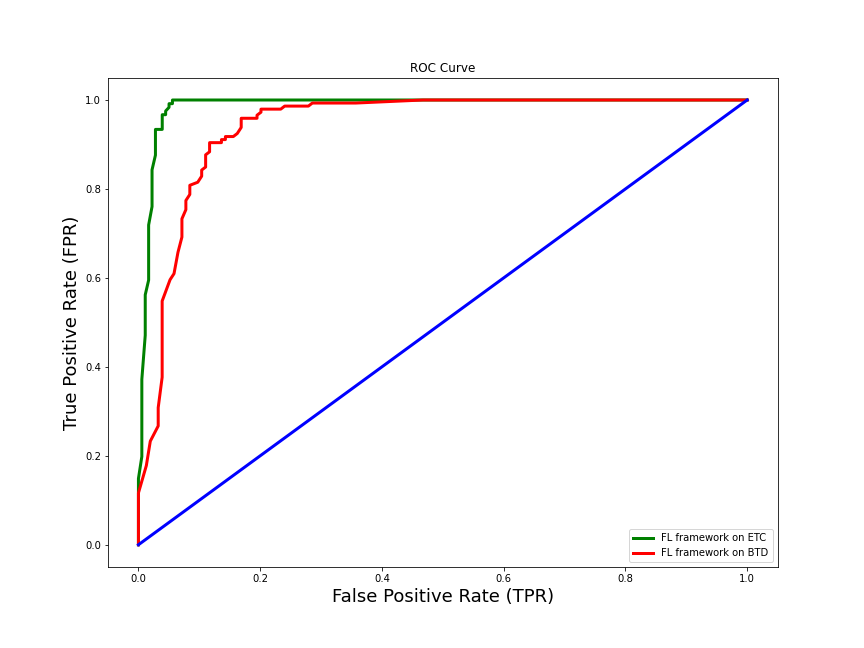}

    \tiny (e) The AUC of ROC curves for NED
  \end{minipage}
  \caption{AUC of ROC curves in FL framework for Block Hunter}
  \label{fig:sample_subfigures1}
\end{figure*}

\begin{comment}
\begin{figure}
    \centering
    \subfigure[The AUC of ROC curves for CBLOF]
    {
        \includegraphics[width=1.9in]{r1.png}
        \label{fig:first_sub1}
    }
    
    \subfigure[The AUC of ROC curves for $K$-means]
    {\includegraphics[width=1.9in]{r2.png}
        \label{fig:second_sub1}
    }
    \subfigure[The AUC of ROC curves for PCA]
    {\includegraphics[width=2.3in]{r3.png}
        \label{fig:third_sub1}
    }
    
    \subfigure[The AUC of ROC curves for IF]
    {\includegraphics[width=2.4in]{r4.png}
        \label{fig:forth_sub1}
    }

        \subfigure[The AUC of ROC curves for NED]
    {
        \includegraphics[width=2.4in]{r5.png}
        \label{fig:fifth}
    }  \caption{AUC of ROC curves in FL framework for Block Hunter }
    \label{fig:sample_subfigures1}
\end{figure}
\end{comment}

\subsubsection{nomaly detection rate}
This subsection aims to assess some well-known machine learning models such as $K$-means, PCA, CBLOF, IF, and NED to hunt anomalies in the Block Hunter framework. We evaluate these models by comparing their average performance, such as Accuracy, Precision, Recall, and F1-score as follows.
%\begin{itemize}
%\item \textbf{	True Positive (TP):} Anomalies correctly identified.
%\item \textbf{	True Negative (TN):} The proportion of normal samples correctly categorized.
%\item \textbf{	False Positive (FP):} Anomalies incorrectly classified as normal samples.
%\item \textbf{	False Negative (FN):} Anomaly examples are incorrectly classified as normal.
%\end{itemize}
These include, Accuracy (Acc) $=\frac{TP + TN}{TP + TN+ FN+FP}$, Recall (Rec) $ =\frac{TP}{TP + FP}$, Precision (Pre) $ = \frac{TP}{TP + FP}$ and F1-score (F1) $ = \frac{2 * TP}{2 * TP+ FN+FP}$.

%according to Equations \ref{eq5}, \ref{eq6}, \ref{eq7}, and \ref{eq8}.
%\begin{equation}\label{eq5}
%\text{Accuracy  (Acc)}=\frac{TP + TN}{TP + TN+ FN+FP}
%\end{equation}
%\begin{equation}\label{eq6}
%\text{Recall  (Rec)} =\frac{TP}{TP + FP}
%\end{equation}
%\begin{equation}\label{eq7}
%\text{Precision  (Pre)}=\frac{TP}{TP + FP}
%\end{equation}
%\begin{equation}\label{eq8}
%\text{F1-score  (F1)}=\frac{2 * TP}{2 * TP+ FN+FP}
%\end{equation}
   
Table \ref{t2} and \ref{t3} display the measured performance of the Block Hunter during applying ML models, $K$-means, PCA, CBLOF, IF, and NED in terms of Precision, Accuracy, F1-score, and Recall based on Bitcoin Transaction and Ethereum Classic Blockchain datasets. To minimize the loss function, all hyperparameters are maximized.  
%
%\begin{figure}
%    \centering
%    \subfigure[The AUC of ROC curves for NED]
 %   {
%        \includegraphics[width=2.2in]{r5.png}
 %       \label{fig:first_sub2}
 %%   }
    
  %  \subfigure[Anomaly detection in the proposed federated framework, $K = 30$, Existing anomaly = 2, Detected = 2]
%    {\includegraphics[width=2.2in]{t1.jpg}
  %      \label{fig:second_sub2}
 %   }\hfill
 %   \subfigure[Anomaly detection in the proposed federated framework, $K = 40$, Existing anomaly = 4, Detected = 4]
 %   {\includegraphics[width=2.2in]{t2.jpg}
 %       \label{fig:third_sub2}
 %   }
    
 %   \subfigure[Anomaly detection in the proposed federated framework, $K = 50$, Existing anomaly = 4, Detected = 4]
  %  {\includegraphics[width=2.2in]{t3.jpg}
  %      \label{fig:forth_sub2}
  %  }
 %   \caption{Anomaly detection in the proposed federated framework}
  %  \label{fig:sample_subfigures}
%\end{figure}

\begin{table}
\tiny
\centering
	\caption{Performance comparison of the ML models based on Bitcoin Transaction Dataset (BTD) }
\begin{tabular}{|c|c|c|c|c|}
\hline
\textbf{Model} & \textbf{Acc (\%)} & \textbf{Pre  (\%) } & \textbf{F1 (\%)} & \textbf{Rec (\%)} \\ \hline
NED            & 96.7\%              & 70.4\%                  & 62.5\%                & 80.2\%              \\ \hline
CBLOF          & 82.4\%              & 55.5\%                 & 78\%                & 66.1\%              \\ \hline
$K$-means        & 87.6\%              & 55.2\%                 & 65.2\%                & 89.5\%              \\ \hline
PCA            & 89.4\%              & 64.2\%                 & 76.2\%               & 76.2\%              \\ \hline
IF             & 95.3\%              & 65.1\%                 & 61.2\%                & 75\%              \\ \hline
\end{tabular}
\label{t2}
\end{table}

\begin{table}[]
\tiny
\centering
	\caption{Performance comparison of the ML models based on Ethereum Classic Blockchain (ETC) }
\begin{tabular}{|c|c|c|c|c|}
\hline
\textbf{Model} & \textbf{Acc (\%)} & \textbf{Pre (\%)} & \textbf{F1 (\%)} & \textbf{Rec (\%)} \\ \hline
NED            & 97.8\%              & 74\%                 & 66.2\%                & 86.2\%              \\ \hline
CBLOF          & 85.6\%              & 60.2\%                 & 82.1\%                & 72.1\%              \\ \hline
$K$-means        & 89.7\%              & 59.1\%                 & 70.2\%                & 93.1\%              \\ \hline
PCA            & 91.6\%              & 70.3\%                 & 81.3\%                & 82.1\%              \\ \hline
IF             & 96.8\%              & 70.5\%                 & 67.6\%                & 81.1v              \\ \hline
\end{tabular}
\label{t3}
\end{table}

In Tables \ref{t2} and \ref{t3}, we can see that NED and IF have the highest accuracy during the anomaly detection while their accuracy is almost similar. In addition, we reported the Area Under the Curve (AUC) of Receiver Operating Characteristic (ROC) as shown in Figs. 4(a), 4(b), 4(c), 4(d), and 4(e). The ROC curves for CBLOF, $K$-means, PCA, IF, and NED are presented in a federated setting. 

By examining the visuals and using the highest level of accuracy metric,the AUC for ROC curves show a comparable ROC curve for all algorithms. The AUC for CBLOF, $K$-means, PCA, IF, and NED are $(0.80, 0.84)$, $(0.82, 0.85)$, $(0.86, 0.89)$, $(0.90, 0.93)$, and $(0.95, 0.97)$  based on BTD and ETC datasets, respectively.
While running the Block Hunter framework with each ML model, we obtain a global model whose parameters are frequently updated via the \texttt{FedAvg} approach \cite{22}. Table   \ref{t6} presents the hunting of anomalies in global models using NED as the local model. This table shows the moment where the Block Hunter framework can hunt an anomaly while doing transactions. This consists of $K = 30, 40, 50$ clusters and $1$ to $35$ transactions per second for $100$ seconds.
Based on the cluster-based structure in Block Hunter, it is almost certain that this system's accuracy is acceptable during anomaly hunting, as shown in Table \ref{t6}. The Block Hunter framework also works perfectly as the number of transactions and clusters increases.

\begin{table}[!ht]
\tiny
\centering
	\caption{Anomaly hunting in Block Hunter}
\begin{tabular}{|c|c|c|c|}
\hline
\textbf{\begin{tabular}[c]{@{}c@{}}Number of cluster \\ (K)\end{tabular}} & \textbf{Existing anomaly} & \textbf{\begin{tabular}[c]{@{}c@{}}Max number of transactions \\  ( per second )\end{tabular}} & \textbf{Detected} \\ \hline
30                                                                        & 2                         & 35                                                                                             & 2                \\ \hline
40                                                                        & 4                         & 35                                                                                           & 4                 \\ \hline
50                                                                        & 4                         & 35                                                                                             & 4                 \\ \hline
30                                                                        & 5                         & 70                                                                                            & 4                 \\ \hline
40                                                                        & 4                         & 70                                                                                            & 3                 \\ \hline
50                                                                        & 4                         & 70                                                                                            & 4                 \\ \hline

\end{tabular}
\label{t6}
\end{table}
{\textcolor{black}{We also evaluated the performance of Block hunter on several IIoT standard datasets as shown in Table  \ref{tf}.  The model performance was evaluated using different ML models namely $K$-means, PCA, CBLOF, IF, and NED on GP and SWaT datasets. NED has the highest accuracy as it preserves data encoding/decoding.}

% \usepackage{multirow}
% \usepackage{multirow}
\begin{table}[]
	\centering
	\caption{\color{black}{Summary of performance comparison of the Block hunter in IIoT related datasets}}
	\tiny
\begin{tabular}{|c|c|c|c|c|c|}
\hline
\textbf{Datasets}     & \textbf{Model} & \textbf{Acc (\%)} & \textbf{Pre (\%)} & \textbf{F1 (\%)} & \textbf{Rec (\%)} \\ \hline
\multirow{5}{*}{GP}   & NED            & 99.1\%            & 98\%              & 99\%             & 98\%              \\ \cline{2-6} 
                      & CBLOF          & 87.8\%            & 82\%              & 78\%             & 70\%              \\ \cline{2-6} 
                      & K -means       & 89.5\%            & 88\%              & 90\%             & 87\%              \\ \cline{2-6} 
                      & PCA            & 95.2\%            & 90\%              & 88\%             & 89\%              \\ \cline{2-6} 
                      & IF             & 96.8\%            & 93\%              & 94\%             & 91\%              \\ \hline
\multirow{5}{*}{SWaT} & NED            & 98.8\%            & 97\%              & 99\%             & 99\%              \\ \cline{2-6} 
                      & CBLOF          & 85.9\%            & 77\%              & 78\%             & 75\%              \\ \cline{2-6} 
                      & K -means       & 87.1\%            & 78\%              & 80\%             & 76\%              \\ \cline{2-6} 
                      & PCA            & 92.1\%            & 89\%              & 88\%             & 90\%              \\ \cline{2-6} 
                      & IF             & 94.6\%            & 95\%              & 90\%             & 92\%              \\ \hline
\end{tabular}
	\label{tf}
\end{table}

%\subsection{The importance of Block hunter}

{\textcolor{black}{Blockchain-based IIoT networks are the underlying technology for the future smart factories, hence an emerging attack target, which shows the significance of this work. To the best of our
knowledge, Block Hunter is the first federated threat hunting
model in IIoT networks that identifies anomalous behavior while preserving privacy. We used FL to build a threat hunting framework that utilizes a cluster-based architecture to formulate an anomaly detection combined with several machine learning models. Our results indicate the superior performance of our model in automatically hunting for anomalies while preserving data privacy.
}}

\section{Conclusion \& Future Works}\label{s6}

In this paper, we developed the Block Hunter framework to hunt anomalies in blockchain-based IIoT smart factories using a federated learning approach. Block Hunter uses a cluster-based architecture to reduce resources and improve the throughput of blockchain-based IIoT networks hunting. The Block Hunter framework was evaluated using a variety of machine learning algorithms (NED, IF, CBLOF, $K$-means, PCA) to detect anomalies. We also examined the impacts of block generation interval, block size, and different miners on the performance of the Block Hunter. Using generative adversarial networks (GAN) to design and implement a block hunter-like framework would be an interesting future research work. Furthermore, \textcolor{black}{ designing and applying IIoT-related blockchain networks with different} consensus algorithms would also be worth investigating in the future.

\bibliographystyle{IEEEtran}
\bibliography{mybibfile15}

\begin{thebibliography}{10}
\providecommand{\url}[1]{#1}
\csname url@samestyle\endcsname
\providecommand{\newblock}{\relax}
\providecommand{\bibinfo}[2]{#2}
\providecommand{\BIBentrySTDinterwordspacing}{\spaceskip=0pt\relax}
\providecommand{\BIBentryALTinterwordstretchfactor}{4}
\providecommand{\BIBentryALTinterwordspacing}{\spaceskip=\fontdimen2\font plus
\BIBentryALTinterwordstretchfactor\fontdimen3\font minus
  \fontdimen4\font\relax}
\providecommand{\BIBforeignlanguage}[2]{{%
\expandafter\ifx\csname l@#1\endcsname\relax
\typeout{** WARNING: IEEEtran.bst: No hyphenation pattern has been}%
\typeout{** loaded for the language `#1'. Using the pattern for}%
\typeout{** the default language instead.}%
\else
\language=\csname l@#1\endcsname
\fi
#2}}
\providecommand{\BIBdecl}{\relax}
\BIBdecl

\bibitem{nn}
J.~Wan, J.~Li, M.~Imran, D.~Li, and F.~e~Amin, ``A blockchain-based solution
  for enhancing security and privacy in smart factory,'' \emph{IEEE
  Transactions on Industrial Informatics}, vol.~15, no.~6, pp. 3652--3660,
  2019.

\bibitem{5i}
F.~Scicchitano, A.~Liguori, M.~Guarascio, E.~Ritacco, and G.~Manco,
  ``Blockchain attack discovery via anomaly detection,'' \emph{Consiglio
  Nazionale delle Ricerche, Istituto di Calcolo e Reti ad Alte Prestazioni
  (ICAR), 2019}, 2019.

\bibitem{4}
Q.~Xu, Z.~He, Z.~Li, M.~Xiao, R.~S.~M. Goh, and Y.~Li, ``An effective
  blockchain-based, decentralized application for smart building system
  management,'' in \emph{Real-Time Data Analytics for Large Scale Sensor
  Data}.\hskip 1em plus 0.5em minus 0.4em\relax Elsevier, 2020, pp. 157--181.

\bibitem{6}
B.~Podgorelec, M.~Turkanovi{\'c}, and S.~Karakati{\v{c}}, ``A machine
  learning-based method for automated blockchain transaction signing including
  personalized anomaly detection,'' \emph{Sensors}, vol.~20, no.~1, p. 147,
  2020.

\bibitem{55i}
\BIBentryALTinterwordspacing
A.~Quintal, ``Veriblock foundation discloses mess vulnerability in ethereum
  classic blockchain,'' \emph{VeriBlock Foundation}. [Online]. Available:
  \url{https://www.prnewswire.com/news-releases/veriblock-foundation-discloses-mess-vulner\\ability-in-ethereum-classic-blockchain-301327998.html}
\BIBentrySTDinterwordspacing

\bibitem{7}
M.~Saad, J.~Spaulding, L.~Njilla, C.~Kamhoua, S.~Shetty, D.~Nyang, and
  D.~Mohaisen, ``Exploring the attack surface of blockchain: A comprehensive
  survey,'' \emph{IEEE Communications Surveys \& Tutorials}, vol.~22, no.~3,
  pp. 1977--2008, 2020.

\bibitem{8}
R.~A. Sater and A.~B. Hamza, ``A federated learning approach to anomaly
  detection in smart buildings,'' \emph{arXiv preprint arXiv:2010.10293}, 2020.

\bibitem{10}
O.~Shafiq, ``Anomaly detection in blockchain,'' Master's thesis, Tampere
  University, 2019.

\bibitem{fff}
A.~Yazdinejadna, R.~M. Parizi, A.~Dehghantanha, and H.~Karimipour, ``Federated
  learning for drone authentication,'' \emph{Ad Hoc Networks}, p. 102574, 2021.

\bibitem{9}
D.~Preuveneers, V.~Rimmer, I.~Tsingenopoulos, J.~Spooren, W.~Joosen, and
  E.~Ilie-Zudor, ``Chained anomaly detection models for federated learning: An
  intrusion detection case study,'' \emph{Applied Sciences}, vol.~8, no.~12, p.
  2663, 2018.

\bibitem{n1}
L.~Tan, H.~Xiao, K.~Yu, M.~Aloqaily, and Y.~Jararweh, ``A blockchain-empowered
  crowdsourcing system for 5g-enabled smart cities,'' \emph{Computer Standards
  \& Interfaces}, vol.~76, p. 103517, 2021.

\bibitem{n2}
L.~Tseng, X.~Yao, S.~Otoum, M.~Aloqaily, and Y.~Jararweh, ``Blockchain-based
  database in an iot environment: challenges, opportunities, and analysis,''
  \emph{Cluster Computing}, vol.~23, no.~3, pp. 2151--2165, 2020.

\bibitem{12}
M.~Signorini, M.~Pontecorvi, W.~Kanoun, and R.~Di~Pietro, ``Bad: a blockchain
  anomaly detection solution,'' \emph{IEEE Access}, vol.~8, pp.
  173\,481--173\,490, 2020.

\bibitem{13}
S.~Iyer, S.~Thakur, M.~Dixit, R.~Katkam, A.~Agrawal, and F.~Kazi, ``Blockchain
  and anomaly detection based monitoring system for enforcing wastewater
  reuse,'' in \emph{2019 10th International Conference on Computing,
  Communication and Networking Technologies (ICCCNT)}.\hskip 1em plus 0.5em
  minus 0.4em\relax IEEE, 2019, pp. 1--7.

\bibitem{14}
S.~Sayadi, S.~B. Rejeb, and Z.~Choukair, ``Anomaly detection model over
  blockchain electronic transactions,'' in \emph{2019 15th International
  Wireless Communications \& Mobile Computing Conference (IWCMC)}.\hskip 1em
  plus 0.5em minus 0.4em\relax IEEE, 2019, pp. 895--900.

\bibitem{16}
Z.~Il-Agure, B.~Attallah, and Y.-K. Chang, ``The semantics of anomalies in iot
  integrated blockchain network,'' in \emph{2019 Sixth HCT Information
  Technology Trends (ITT)}.\hskip 1em plus 0.5em minus 0.4em\relax IEEE, 2019,
  pp. 144--146.

\bibitem{17}
F.~Scicchitano, A.~Liguori, M.~Guarascio, E.~Ritacco, and G.~Manco, ``A deep
  learning approach for detecting security attacks on blockchain.'' in
  \emph{ITASEC}, 2020, pp. 212--222.

\bibitem{w1}
T.~R. Gadekallu, Q.-V. Pham, D.~C. Nguyen, P.~K.~R. Maddikunta, N.~Deepa,
  B.~Prabadevi, P.~N. Pathirana, J.~Zhao, and W.-J. Hwang, ``Blockchain for
  edge of things: Applications, opportunities, and challenges,'' \emph{IEEE
  Internet of Things Journal}, vol.~9, no.~2, pp. 964--988, 2022.

\bibitem{18}
T.~D. Nguyen, S.~Marchal, M.~Miettinen, H.~Fereidooni, N.~Asokan, and A.-R.
  Sadeghi, ``D{\"i}ot: A federated self-learning anomaly detection system for
  iot,'' in \emph{2019 IEEE 39th International Conference on Distributed
  Computing Systems (ICDCS)}.\hskip 1em plus 0.5em minus 0.4em\relax IEEE,
  2019, pp. 756--767.

\bibitem{19}
S.~Li, Y.~Cheng, Y.~Liu, W.~Wang, and T.~Chen, ``Abnormal client behavior
  detection in federated learning,'' \emph{arXiv preprint arXiv:1910.09933},
  2019.

\bibitem{20}
R.~Kumar, A.~A. Khan, S.~Zhang, W.~Wang, Y.~Abuidris, W.~Amin, and J.~Kumar,
  ``Blockchain-federated-learning and deep learning models for covid-19
  detection using ct imaging,'' \emph{arXiv preprint arXiv:2007.06537}, 2020.

\bibitem{21}
H.~Chai, S.~Leng, Y.~Chen, and K.~Zhang, ``A hierarchical blockchain-enabled
  federated learning algorithm for knowledge sharing in internet of vehicles,''
  \emph{IEEE Transactions on Intelligent Transportation Systems}, 2020.

\bibitem{w2}
M.~Alazab, S.~P. RM, P.~M, P.~K.~R. Maddikunta, T.~R. Gadekallu, and Q.-V.
  Pham, ``Federated learning for cybersecurity: Concepts, challenges, and
  future directions,'' \emph{IEEE Transactions on Industrial Informatics},
  vol.~18, no.~5, pp. 3501--3509, 2022.

\bibitem{22}
H.~Brendan~McMahan, E.~Moore, D.~Ramage, S.~Hampson, and B.~Ag{\"u}era~y Arcas,
  ``Communication-efficient learning of deep networks from decentralized
  data,'' \emph{ArXiv e-prints}, pp. arXiv--1602, 2016.

\bibitem{23}
J.~Kone{\v{c}}n{\`y}, H.~B. McMahan, F.~X. Yu, P.~Richt{\'a}rik, A.~T. Suresh,
  and D.~Bacon, ``Federated learning: Strategies for improving communication
  efficiency,'' \emph{arXiv preprint arXiv:1610.05492}, 2016.

\bibitem{24}
N.~Moussa and A.~E.~B. El~Alaoui, ``An energy-efficient cluster-based routing
  protocol using unequal clustering and improved aco techniques for wsns,''
  \emph{Peer-to-Peer Networking and Applications}, pp. 1--14, 2021.

\bibitem{a4}
A.~Yazdinejad, R.~M. Parizi, G.~Srivastava, A.~Dehghantanha, and K.-K.~R. Choo,
  ``Energy efficient decentralized authentication in internet of underwater
  things using blockchain,'' in \emph{2019 IEEE Globecom Workshops (GC
  Wkshps)}.\hskip 1em plus 0.5em minus 0.4em\relax IEEE, 2019, pp. 1--6.

\bibitem{27}
V.~Le, T.~P. Quinn, T.~Tran, and S.~Venkatesh, ``Deep in the bowel: highly
  interpretable neural encoder-decoder networks predict gut metabolites from
  gut microbiome,'' \emph{BMC genomics}, vol.~21, no.~4, pp. 1--15, 2020.

\bibitem{28}
S.~Golovkine, N.~Klutchnikoff, and V.~Patilea, ``Clustering multivariate
  functional data using unsupervised binary trees,'' \emph{arXiv preprint
  arXiv:2012.05973}, 2020.

\bibitem{30}
H.~Abdi and L.~J. Williams, ``Principal component analysis,'' \emph{Wiley
  interdisciplinary reviews: computational statistics}, vol.~2, no.~4, pp.
  433--459, 2010.

\bibitem{31}
K.~P. Sinaga and M.-S. Yang, ``Unsupervised k-means clustering algorithm,''
  \emph{IEEE Access}, vol.~8, pp. 80\,716--80\,727, 2020.

\bibitem{gas}
I.~P. Turnipseed, \emph{A new scada dataset for intrusion detection
  research}.\hskip 1em plus 0.5em minus 0.4em\relax Mississippi State
  University, 2015.

\bibitem{sw}
R.~Taormina, S.~Galelli, N.~O. Tippenhauer, E.~Salomons, A.~Ostfeld, D.~G.
  Eliades, M.~Aghashahi, R.~Sundararajan, M.~Pourahmadi, M.~K. Banks
  \emph{et~al.}, ``Battle of the attack detection algorithms: Disclosing cyber
  attacks on water distribution networks,'' \emph{Journal of Water Resources
  Planning and Management}, vol. 144, no.~8, p. 04018048, 2018.

\end{thebibliography}

\end{document}